\documentclass[smallextended,twocolumn]{svjour3}
\smartqed
\usepackage{graphicx}
\usepackage{color}
\usepackage{ulem}
\usepackage{amsmath}
\usepackage{amssymb}

\newcommand{\be}{\begin{equation}}
\newcommand{\ee}{\end{equation}}

\newcommand{\bex}{\begin{eqnarray}}
\newcommand{\eex}{\end{eqnarray}}
\newcommand{\bmin}{\begin{center}\begin{minipage}{460pt}}
\newcommand{\emin}{\end{minipage}\end{center}}

\begin{document}

\title{The essence of nonclassicality: non-vanishing signal deficit}

\author{S. Aravinda \and R. Srikanth } 

\institute{S.  Aravinda  \at   Poornaprajna  Institute  of  Scientific
  Research,  \\ Bangalore,  India.  \and R.  Srikanth \at  Poornaprajna
  Institute  of  Scientific Research,  \\  Bangalore,  India.}

\date{Received:date / Accepted: date}

\maketitle

\begin{abstract}
Nonclassical  properties  of   correlations--  like  unpredictability,
no-cloning and uncertainty-- are known to follow from two assumptions:
nonlocality and no-signaling.   For two-input-two-output correlations,
we derive these properties from  a single, unified assumption: namely,
the  excess  of the  communication  cost  over  the signaling  in  the
correlation.   This  is  relevant to  quantum  temporal  correlations,
resources to  simulate quantum correlations and  extensions of quantum
mechanics.   We generalize  in the  context of  such correlations  the
nonclassicality   result    for   nonlocal-nonsignaling   correlations
(Masanes,  Acin  and  Gisin,  2006)   and  the  uncertainty  bound  on
nonlocality  (Oppenheim  and  Wehner,  2010),  when  the  no-signaling
condition  is relaxed.
\end{abstract}

\keywords{Bell's theorem \and nonclassicality \and signaling}

\section{Introduction \label{intro}}

It  is known  that bi-partite  correlations in  nonclassical theories,
such as quantum  mechanics (QM) or the world of  PR boxes \cite{PR94},
possess common properties absent  in classical probability theory, and
these properties can be traced  to two basic assumptions: no-signaling
and nonlocality  \cite{MAG06}.  By ``nonclassical'', we  mean features
like unpredictability  \cite{cw12} of outcomes of  measurement on pure
states,  uncertainty of  conjugate pairs  of observables,  monogamy of
nonlocal correlations \cite{Ton06}, no-cloning \cite{WZ,Gis}, etc.  By
`pure states', we mean the extreme points of the polytope of the state
space.

In this  work, we  unify these  two assumptions  into a  single weaker
assumption: namely, the existence of a  gap $\eta$ by which the signal
in bipartite  correlation \textbf{P} falls short  of the communication
cost of \textbf{P}.  We describe  correlation \textbf{P} as the vector
$P_{AB|ab}$,  where   $a$  and  $b$   are  Alice's  and   Bob's  input
measurements  respectively,  and $A$  and  $B$  are their  measurement
outcomes.  We  show that the  condition $\eta>0$ suffices  to generate
nonclassical properties \cite{AS2}, even if  we relax the condition of
no-signaling.    Conventionally,  almost   all  research   on  quantum
correlations has been in the  no-signaling scenario, and here we point
to a direction to go beyond  this.  In particular, our results will be
relevant to the study of temporal correlations \cite{DAS+0}, resources
required to simulate QM \cite{ASQIC}, and extensions of QM.

\section{Correlation inequality and signaling}

The  correlations  $\textbf{P}=P_{AB|ab}$  we consider  here  will  be
restricted to measurement outcomes $A,B=\pm 1$ with measurement inputs
$a,b = {0,1}$. \textbf{P} fails  to admit a deterministic local hidden
variable (LHV) model \cite{fine82}:
\begin{equation}
P_{AB|ab}   =    \int\rho(\lambda)P(A|a,\lambda) P(B|b,\lambda) d\lambda
\label{eq:HV}
\end{equation}
if it violates the condition
\begin{equation}
\Lambda(\textbf{P}) \equiv  |E(0,0) + E(0,1)  - E(1,0) +
E(1,1)| \leq 2,
\label{eq:corr}
\end{equation}
where $E(a,b)  = \sum_{A,B} AB\times P_{AB|ab}$  indicates the average
outcome upon  measuring two  given input observables,  and $P_{AB|ab}$
satisfies  the positivity and  normalization conditions.   By checking
the  non-violation  of  each  of  the CHSH  inequalities  obtained  by
permuting settings  and outcomes  (which here effectively  gives three
other  inequalities,  with the  minus  sign  in Ineq.  (\ref{eq:corr})
displaced), we would know that \textbf{P} admits a LHV model.

If  $(a=0,  a=1)$  and  $(b=0,  b=1)$  refer  to  spatially  separated
mesurements  on  two  different  particles, then  \textbf{P}  must  be
non-signaling and  Eq.  (\ref{eq:HV}) corresponds to  the condition of
\textit{local-realism}   and  Ineq.    (\ref{eq:corr})   is  the   the
Clauser-Horne-Shimony-Holt (CHSH) inequality  \cite{CHSH}, a Bell-type
inequality \cite{Bel64}.  By contrast, if $(a=0,a=1)$ and $(b=0, b=1)$
refer to temporally separated measurements  on the same particle, then
Eq.     (\ref{eq:HV})     corresponds    to    the     condition    of
\textit{noninvasive-realism}  and Ineq.   (\ref{eq:corr})  is the  the
Leggett-Garg  (LG)  inequality  \cite{lg85} in  its  two-time  variant
\cite{bruk04}. Unlike in the spatial  case, relativity does not forbid
\textbf{P}  from   being  signaling   in  the  temporal   case.  Since
Eq. (\ref{eq:HV}) implies the no-signaling condition
\begin{equation} 
\sum_{B}P_{AB|00}=\sum_{B}P_{AB|01};~~
\sum_{A}P_{AB|00}=\sum_{A}P_{AB|10},
\label{eq:nosig}
\end{equation}
the presence of signal in \textbf{P} by itself guarantees violation of
noninvasive-realism.    The   situation   with   \textit{contextuality
  inequalities}  is similar  to Bell-type  inequalities, in  that they
must be non-signaling \cite{KCB+08}.

As  our results  below  will apply  to  any of  these  three kinds  of
inequalities, it will  be convenient to use a  uniform terminology and
refer to  (\ref{eq:corr}) as a \textit{correlation  inequality} and to
Eq.  (\ref{eq:HV}) as the \textit{separability} condition.

\section{The signaling polytope and communication cost}

It  will  be  convenient  to  characterize  a  bi-partite  correlation
\textbf{P} in terms of two parameters, the signaling and communication
cost.  One way to define the signal from Alice to Bob is by
\begin{equation}
s_{\mathcal{A}\rightarrow \mathcal{B}}(\textbf{P}) =
\max_{b}|P(B=0|0,b) - P(B=0|1,b)|,
\label{defn:signal}
\end{equation}
and  similarly  for  the  signal   from  Bob  to  Alice.   The  signal
$s(\textbf{P})$   is   the   maximum   of   $s_{\mathcal{A}\rightarrow
  \mathcal{B}}$ and $s_{\mathcal{B}\rightarrow \mathcal{A}}$.

The average communication cost $C$ is  the minimum number of bits that
Alice must send  to Bob in a classical simulatation  of \textbf{P}. In
general, $C$ must convey something  to Bob about both Alice's settings
and outcomes \cite{PKP+10}, but assuming  freewill of both players and
allowing for outcome information to be part of pre-shared information,
$C$ only carries her settings information, $a$.

The   bi-partite  2-input-2-output   possibly   signaling  correlation
\textbf{P}, which has  4 possible inputs and 4  possible outputs, is a
list  of  $4 \times  4$  numbers.   Taking  into consideration  the  4
probability  conservation conditions  for each  input, but  not  the 4
independent  no-signaling conditions  (\ref{eq:nosig}),  there are  12
free  parameters,  which  is  the  dimension  $D_\mathcal{S}$  of  the
``signaling polytope'' $\mathcal{S}$.  The  4 possible outputs on each
input entails that there are  $4^4 = 256$ deterministic \textbf{P}, or
deterministic  boxes  \textbf{d},  that  are  the  extreme  points  of
$\mathcal{S}$.   This is  appropriate  for a  classical simulation  of
\textbf{P} \cite{Pir03}.

Of the \textbf{d}'s, sixteen  are deterministic 0-bit boxes (for which
$C=0$),  and are  the extreme  points of  the  \textit{local polytope}
$\mathcal{L}$, and the remaining 240 are deterministic boxes requiring
1 bit  (in the case of  1-way signaling) or  2 bits (in case  of 2-way
signaling)   for   their   simulation  \cite{MAS12}.    The   familiar
no-signaling polytope $\mathcal{N}$ is  a subset of $\mathcal{S}$, and
exists in an  8-dimensional space. It has 24 pure  states, 16 of which
are the pure  points of $\mathcal{L}$, while the  remaining 8 of which
are the PR boxes \cite{PR94}.

A (convex) polytope  can be defined in terms of  its vertices or facet
inequalities.  It  turns out  that Ineq.   (\ref{eq:corr}) is  a facet
inequality for the local polytope  \cite{Pir03}.  Eight of these local
deterministic boxes are:
\begin{equation}
\begin{tabular}{c||c|c|c|c|c|c|c|c}
\hline  $ab$  &  $\textbf{d}^{0_0}$  &  $\textbf{d}^{1_0}$ &  
$\textbf{d}^{2_0}$  &  $\textbf{d}^{3_0}$  &
$\textbf{d}^{4_0}$& $\textbf{d}^{5_0}$ & $\textbf{d}^{6_0}$ & $\textbf{d}^{7_0}$ 
\\ \hline 00 & 00 & 00 & 01 & 11 &  00 & 10 & 11 & 11\\ 
          01 & 00 & 00 & 00 & 10 & 01 & 11 & 11 & 11 \\ 
          10 & 00 & 10 & 01 & 01 & 10 & 10 & 01 & 11 \\ 
          11 & 00 & 10 & 00 & 00 & 11  & 11 & 01 & 11 \\ \hline
\end{tabular}
\label{eq:d0}
\end{equation}
for which  $\Lambda=+2$ in  Ineq.  (\ref{eq:corr}), while  eight 1-way
signaling \textbf{d}'s are:
\begin{equation} 
\begin{tabular}{c||c|c|c|c|c|c|c|c}
\hline  $ab$  &  $\textbf{d}^{0_1}$ & $\textbf{d}^{1_1}$ &
$\textbf{d}^{2_1}$ & $\textbf{d}^{3_1}$ & $\textbf{d}^{4_1}$  & 
$\textbf{d}^{5_1}$ & $\textbf{d}^{6_1}$ & $\textbf{d}^{7_1}$
\\ \hline 
00 &  00 & 11 & 00 & 11 & 00 & 11 & 00 & 11 \\ 
01 &  00 & 11 & 00 & 11 & 00 & 00 & 11 & 11 \\ 
10 &  01 & 01 & 10 & 10 & 10 & 01 & 10 & 01\\ 
11 &  00 & 00 & 11 & 11 & 00 & 00 & 11 & 11
\\ \hline
\end{tabular}
\label{eq:d1}
\end{equation}
for which $\Lambda=+4$ in Ineq.  (\ref{eq:corr}).  For spatial quantum
correlations, the decomposition in terms of the deterministic boxes of
Eqs.   (\ref{eq:d0}) and  (\ref{eq:d1})  is optimal  to determine  $C$
\cite{Pir03,Pir03+}.

As  the number  of  these different  deterministic  boxes exceeds  the
dimension  $D_{\mathcal{S}}$,  in  general,  there  will  be  multiple
decompositions of \textbf{P}:
\begin{equation} 
\textbf{P}    \equiv     P_{AB|ab}    =    \sum_{\lambda     =    0}^7
q_{\lambda_0}\textbf{d}^{\lambda_0}_{AB|ab}+\sum_{\lambda=0}^7
q_{\lambda_1}\textbf{d}^{\lambda_1}_{AB|ab}.
\label{eq:decomp}
 \end{equation}
Of these, the optimal decomposition is one for which the quantity
\begin{equation}
C(\textbf{P})  = \sum_{\lambda=0}^7 q_{\lambda_1}
\label{eq:CC}
\end{equation}
must be minimum, and quantifies the average communication cost.

Now we  note that $C$ for  a mixture of deterministic  correlations is
simply  the  average of  the  communication  cost of  the  constituent
deterministic boxes. For example
\begin{equation}
C\left(p\textbf{d}^{0_1}   +
q\textbf{d}^{2_1}\right)= p+q.
\label{eq:Cav}
\end{equation}
whereas the signaling  $s$ is a convex function.  This  is because the
pattern of signaling, namely Bob's  input to observe Alice's signal or
Bob's output  for his  same input,  will not be  the same  between two
1-bit boxes.  For  example, in the determinsitic 1-bit  boxes of Table
\ref{eq:d1},   to   receive   Alice's   signal,   in   the   case   of
$\textbf{d}^{0_1}$, Bob  chooses $b=0$, while  for $\textbf{d}^{2_1}$,
he chooses $b=1$.

One finds that
\begin{eqnarray}
s\left(p\textbf{d}^{0_1} + q\textbf{d}^{2_1}\right)&=& \max(p,q) \le p+q.
\nonumber \\
s\left(p\textbf{d}^{0_1}    +    q\textbf{d}^{3_1}\right)
&=& |p-q| \le  p+q.
\label{eq:unfriendly}
\end{eqnarray}
It  may be  verified that  this bound holds for  any pair  of 1-bit
strategies (\ref{eq:d1}), and thus we have:
\begin{equation}
s(\textbf{P}) \le \sum_{\lambda}^7 q_{\lambda_1}.
\label{eq:signal}
\end{equation}
Combining Eqs (\ref{eq:CC}) and (\ref{eq:signal}) we have:
\begin{equation}
s(\textbf{P})  \leq C(\textbf{P}).  
\label{eq:metamath}
\end{equation} 
A more general proof of Ineq.  (\ref{eq:metamath}) for \textbf{P} with
arbitrary number of inputs and outputs, using an entropic argument, is
presented elsewhere \cite{ASX}.

\section{Geometric nonclassicality}

Now  suppose   a  pair   of  1-bit  strategies   (\ref{eq:d1}),  e.g.,
$(\textbf{d}^{0_1}$  and $\textbf{d}^{3_1})$,  occurs together  in the
optimal    decomposition     (\ref{eq:decomp}),    then    necessarily
$s(\textbf{P})  < C(\textbf{P})$  because of  the different  geometric
properties of $s$ and $C$, as  discussed above.  Thus, mixing any such
pair  of  strategies  results  in  the  gap  $\eta(\textbf{P})  \equiv
C(\textbf{P}) - s(\textbf{P})$  being larger than 0.   The converse is
also true,  i.e., if $\eta(\textbf{P})$  is larger than 0,  then there
are  two or  more  distinct 1-bit  strategies  occuring with  non-zero
probabibility in the optimal  decomposition (\ref{eq:decomp}).  To see
this,  note that  in Eq.   (\ref{eq:decomp})  only if  a single  1-bit
strategy occurs is it the case that $s(\textbf{P}) = C(\textbf{P})$.

In  other words,  if $\eta  >0$, then  there is  \textit{non-separable
  unpredictability}-- i.e.,  mixing of more than  one 1-bit strategies
of the type (\ref{eq:d1}) with non-zero probability.  Note that mixing
0-bit strategies like (\ref{eq:d0})  cannot lead to $\eta>0$.  Without
distinguishing between  the non-separable and local  contributions, we
may quantify local unpredictability by:
\begin{equation}
I(\textbf{P}) \equiv 
\max_{a,b} \min_{o}\{P_{o|a,b}, 1-P_{o|a,b}\},
\label{eq:IR}
\end{equation}
where $o$ is $A$ or $B$.  Here $I(\textbf{P})$ is so called because it
is obtained by considering the  unpredictability observed by Alice and
Bob  locally, and  then  taking  the larger  of  them.  If  \textbf{P}
corresponds to a pure state, then $I(\textbf{P})>0$ says that there is
fundamental unpredictability in the theory.

Whereas  the results  so far  are quite  general, still  what we  call
classical or otherwise is a matter of (aesthetic) choice.  We indicate
three criteria based on unpredictability:
\begin{description}
\item[{\bf  C0}.]   According  to  the most  stringent  definition  of
  classicality,  classical pure  states  necessarily give  predictable
  outcomes.   Correspondingly, we  get the  weakest interpretation  of
  nonclassicality:  any theory  with pure  state having  non-vanishing
  $I(\textbf{P})$.   {\it   Separable-state  QM}  (which   is  the
  fragment of QM whose state  space is restricted to separable states,
  and   whose  allowed   operations  preserve   the  separable   state
  structure) is ``weakly nonclassical'',  because of randomness in
  the outcomes  of measuring  states that are  not eigenstates  of the
  measured  observable.  Further,  since $\eta(\textbf{P})>0$  implies
  $I(\textbf{P})>0$,  thus signal  deficit  states  (i.e., those  with
  $s<C$) are nonclassical.
\item[{\bf  C1}.]   A  weaker   definition  of  classicality,  whereby
  classical  pure  states are  any  states  whose measurement  outcome
  statistics  can be  simulated  by deterministic  non-contextual
  hidden variables.   Separable-state QM  then is  nonclassical by
  this criterion, as proven by the Kochen-Specker theorem \cite{KS67}.
\item[{\bf  C2}.]  A  still weaker  definition of  classicality,
  defined as  characterizing a  theory whose state  space lies  in the
  local polytope, where measurement outcomes can be simulated by local
  \textit{indeterministic}     non-contextual    hidden     variables.
  Seperable-state QM is \textit{classical}  by this criterion, in that
  there is an indeterministic but measurement non-contextual model for
  separable-state  QM: the  Beltrametti-Bugajski model  \cite{BB95} as
  applied to the separable-state QM.  Correspondingly one has the most
  stringent interpretation of nonclassicality.  Even so, a theory that
  allows $\eta>0$ is ``strongly  nonclassical'', because even allowing
  for  outcome indeterminism,  a local  hidden variable  theory cannot
  model states with $\eta>0$.
 \end{description} 

It is  worth noting that  the above considerations are  purely formal,
and do not  depend on the physical interpretation  of the correlations
as  being spatial  (the  measurements  of Alice  and  Bob  are on  two
distinct particles) or temporal (the measurements of Alice and Bob are
sequentially on the same particle).  In the former case, Alice and Bob
refer to  observations ``here'' and  ``there'', while in  the temporal
case, they refer to observations  ``now'' and ``later''.  There is one
caveat,   though:  if   the   correlation   \textbf{P}  is   signaling
($s(\textbf{P})>0$),  then  the  chronology of  measurements  must  be
consistent  with  the signal,  with  the  signal sender's  measurement
preceding the receiver's. (Otherwise, the free will of the sender gets
restricted.)

This perspective allows us to study the nonclassicality of spatial and
temporal  correlations on  the same  footing.  Both  the violation  of
Bell's inequality and the violation of the Leggett-Garg inequality are
strongly  nonclassical, by  criterion (C2),  since the  ``here-there''
correlations  relevant  to  the  former inequality,  as  well  as  the
``now-later'' correlations relevant to  the latter inequality, contain
non-separable unpredictability.

If  Alice and  Bob hold  $d$-level  systems, then  the largest  signal
possible by transmitting  such a system is $\log(d)$.   In the context
of   temporal   correlations,   dimension  witnesses   are   Bell-type
inequalities for  successive preparation and measurement  on $d$-level
quantum systems,  whose violation requires  a communication cost  $C >
\log(d)$.  Therefore, irrespective of  the signal level, a correlation
that  violates  a  dimension  witness  \cite{GBH+10}  is  nonclassical
according to  the strongest criterion (C2).   We can in fact  define a
`super-strong' criterion  of nonclassicality  (say, C3),  according to
which   the   necessary   condition  for   nonclassicality   is   that
$\eta^\ast>0$,   where   $\eta^\ast\equiv    C-\log(d)$   (cf.    Ref.
\cite{BKM+15}).  However, such a criterion would be applicable only to
temporal correlations, and not to spatial correlations, and would thus
be  unsuitable  to  our  present  purpose  of  identifying  the  basic
nonclassical elements  of correlations,  that would be  indifferent to
the spatio-temporal status of the correlations.

While  $\eta>0$ implies  nonclassicality,  the converse  is not  true,
since one  can still  have weak non-classicality  according to  the C0
criterion.    We   may   think    of   C2-nonclassicality   as   being
\textit{Bell-certified},   while   some  C0-nonclassicality   can   be
simulated   using   local   randomness.    A   weakly-but-not-strongly
nonclassical theory would be one  that is locally nonclassical, in the
sense that no-cloning holds good in each local sector.  Consider a toy
theory  $\mathfrak{Q}_0$   in  which   the  local   correlations  like
(\ref{eq:d0})  and  1-bit  correlations like  (\ref{eq:d1})  are  pure
states.  Geometrically, the pure-state  decomposition for mixed states
in this theory will not be unique, \textit{even for local states}.  To
see this, we note that:
\begin{eqnarray}
\frac{1}{2}\left(\textbf{d}^{0_0}   +   \textbf{d}^{7_0}\right)  =
\frac{1}{2}\left(\textbf{d}^{3_0} + \textbf{d}^{4_0}\right).
\label{eq:nonuniq}
\end{eqnarray}
Multiciplicity of decomposition implies that  the state space is not a
simplex, and contains a no-cloning  theorem, making it nonclassical in
that sense \cite{BBL+07}.

Finally, a classical theory is  one where pure states lack fundamental
unpredictability,  and  thus  is   classical  even  by  the  strongest
criterion C0. An example of a classical theory would be one with 0-bit
correlation states  like those given  by Eq.  (\ref{eq:d0}),  and with
1-bit correlation states  like those given by  Eq.  (\ref{eq:d1}). But
in  contrast to  $\mathfrak{Q}_0$,  these correlations  are no  longer
`boxes' but  instead are strategies for  \textit{classical simulation}
of  local  or  nonlocal  correlations.  Thus  the  state  space  is  a
255-dimensional  simplex, whose  vertices  are all  the 256  two-party
two-input-two-output  deterministic strategies  like  those listed  in
Eqs.  (\ref{eq:d0}) and (\ref{eq:d1}).  Further, as the state space is
simplex,  this classical  theory lacks  no-cloning.  By  contrast, the
theory $\mathfrak{Q}_0$  is a  non-simplex polytope  in $4\cdot4-4=12$
dimensions.

\section{Signaling undermines nonclassicality \label{sec:noncla}}

The  preceding  Section  shows  that  the  gap  between  $C$  and  $s$
guarantees unpredictability.  This qualitative observation can be made
quantitative,  and can  be extended  to other  nonclassical properties
besides unpredictability \cite{AS2,ASX}, like no-cloning, uncertainty,
monogamy, etc., that are consequences of the assumption of nonlocality
($C>0$) and  no-signaling ($s=0$) \cite{MAG06}.   In the case  of each
property,   one  can   show   that  the   property  persists,   though
diminishingly, when the signal level is raised at constant $C$.

\subsection{Fundamental unpredictability}

Here  we  illustrate  the  idea  quantitatively  for  the  fundamental
unpredictability   and   uncertainty    in   an   operational   theory
$\mathcal{T}$.  Fundamental  unpredictability is  the quantity  in Eq.
(\ref{eq:IR}) maximized over all pure states $\psi$ in $\mathcal{T}$:
\begin{equation}
I(\mathcal{T}) = \sup_\psi I(\textbf{P}_\psi),
\end{equation}
where  $\textbf{P}_\psi$  is  the   correlation  generated  by  making
measurements     on    $\psi$. 

For state $\psi$ characterized by communication cost $C$, there exists
a complementarity  between the signaling  and local randomness  in any
resource (PR  boxes, classical  communication, signaling  boxes, etc.)
that simulates $\psi$ \cite{ASQIC}:
\begin{equation}
s + 2I \ge C,
\label{eq:as3}
\end{equation}
from which it follows by direct substitution that
\begin{equation}
I \ge \frac{\eta}{2}.
\label{eq:Eta0}
\end{equation}
The  result in  Ref. \cite{MAG06},  that fundamental  unpredictability
(referred  to as  ``intrinsic  randomness'' in  that  reference) is  a
consequence of  no-signaling and nonlocality, is  generalized in Ineq.
(\ref{eq:Eta0}) in  the context of  two-input-two-output correlations,
by relaxing the no-signaling condition.

Ineq.    (\ref{eq:Eta0})  can   be   interpretted   as  showing   that
unpredictability  that cannot  be modelled  using separable  resources
(i.e., ``Bell-certified randomness'') is weakened by signaling, in the
sense that for  fixed degree of nonseparability as  quantified by $C$,
an  increase  in $s$  reduces  the  lower  bound  on $I$.   From  this
viewpoint,  signaling can  be said  to undermine  \textit{strong} (C2)
nonclassicality.  One can still  have weak (C0) nonclassicality, i.e.,
unpredictability  originating  from  the separable  component  of  the
correlations.

Suppose   we   are   given  two   correlations,   $\textbf{P}_n$   and
$\textbf{P}_s$,  both  violating  Ineq. (\ref{eq:corr})  at  the  same
level, but  with the  former correlation  being non-signaling  and the
latter  signaling.   Ineq. (\ref{eq:Eta0})  does  not  imply that  the
signaling correlation is  less classical.  This is  because the degree
of  violation  of  a  correlation  inequality  only  lower-bounds  $C$
\cite{Pir03}.   It   may  be   the  case  that   $C(\textbf{P}_s)  \ge
C(\textbf{P}_n)+s$,      so      that     $\eta(\textbf{P}_s)      \ge
\eta(\textbf{P}_n)$,  meaning that  the signaling  correlation can  be
more C2-nonclassical.

\subsection{Uncertainty}

Uncertainty measures the incompatibility of two observables, say $a=0$
and  $a=1$.   We  will  find  it convenient  to  use  the  concept  of
uncertainty  directly related  to unpredictability,  though any  other
definition (entropic, standard deviation or fine-grainded) would do as
well. On any one side, the uncertainty on input may be quantified as:
\begin{eqnarray}
\Delta_\mathcal{A}^a &=& \max_b\min_A P_{A|a,b}, \nonumber\\
\Delta_\mathcal{B}^b &=& \max_a\min_B P_{B|a,b},
\label{eq:Delta}
\end{eqnarray}
which is unpredictability (\ref{eq:IR}),  but without the maximization
over the local input.  Uncertainty exists on Alice's side if
\begin{equation}
\mathcal{U}_{\mathcal{A}} \equiv \Delta_{\mathcal{A}}^0 + \Delta_{\mathcal{A}}^1 > 0,
\label{eq:uncertainty}
\end{equation}
and  similarly  for  Bob.   Inequality  (\ref{eq:uncertainty})  is  an
uncertainty relation because it says that  there is no state such that
both measurements $a=0$ and $a=1$ are simultaneously predictable.

Now consider  a pure state $\psi$  in an operational theory,  which is
given  by a  mixture of  the 0-bit  strategies $\textbf{d}^{0_0}$  and
$\textbf{d}^{2_0}$.  One finds  $\mathcal{U}_{\mathcal{A}}=0$. Now let
us suppose  a non-vanishing gap $\eta$.  This comes from a  mixture of
$\textbf{d}^{j_1}$  strategies, which  for simplicity,  we confine  to
those that signal  from Alice to Bob, i.e., the  first four strategies
of (\ref{eq:d1}), whose equations are given by:
\begin{equation}
\begin{array}{c|l|l}
\hline
\textbf{d}^{0_1} ~&~ A=0 ~&~ B=a\cdot(b+1) \\
\hline 
\textbf{d}^{3_1} ~&~ A=1 ~&~ B=a\cdot(b+1) + 1\\ \hline 
\textbf{d}^{2_1} ~&~ A=a ~&~ B=a\cdot b \\ \hline
\textbf{d}^{1_1} ~&~ A=a+1 ~&~ B=a\cdot b + 1 \\
\hline
\end{array}
\label{eq:d1x}
\end{equation}
where  `+'  indicates  mod-2  addition.   These  four  strategies  are
obtained by imposing the locality condition  (column 2 of the Table in
Eq. (\ref{eq:d1x}))  on the CHSH  condition $x + y=a\cdot  b+1$, which
would  violate  Ineq.   (\ref{eq:corr})   to  its  algebraic  maximum.
Consider a decomposition (\ref{eq:decomp}) whose 1-bit part mixes only
two  of  these,  say $\textbf{d}^{0_1}$  and  $\textbf{d}^{2_1}$  with
probabilities $p_0$ and $p_2$.

Without   loss   of   generality,   let   $p_0\le   p_2$.    We   find
$\Delta^0_{\mathcal{A}}=0$ while $\Delta^1_{\mathcal{A}}=p_0$, and so,
using                definition                (\ref{eq:uncertainty}),
$\mathcal{U}_{\mathcal{A}}=p_0$.  On the other  hand, $s \ge p_2-p_0 =
C-2p_0$, so that:
\begin{equation}
s + 2\mathcal{U}_{\mathcal{A}} \ge C,
\label{eq:ucmp}
\end{equation}
which is analogous to  the result (\ref{eq:as3}) for unpredictability.
In  general, by  mixing two  distinct 1-bit  boxes with  probabilities
$p_{\min}$   and   $p_{\max}$   with   $p_{\min}<p_{\max}$,   we   get
$\mathcal{U}\ge p_{\min}$ and $s\ge  p_{\max}-p_{\min}$.  Thus we have
$s   +  2\mathcal{U}   \ge   p_{\max}+p_{\min}  =   C$.   The   result
(\ref{eq:ucmp})  can  be  shown  to hold  for  arbitrary  mixtures  of
strategies  separable and  non-separable strategies  (\ref{eq:d0}) and
(\ref{eq:d1}), and for uncertainty of Bob, too \cite{ASX}.

Rewriting Ineq. (\ref{eq:ucmp}), we have:
\begin{equation}
\mathcal{U}_{\mathcal{A}} \ge \frac{\eta}{2}.
\label{eq:U}
\end{equation}
This   generalizes   from   nonsignaling   to   (possibly)   signaling
correlations, in the context of two-input-two-output correlations, the
result  of   \cite{MAG06}  that   uncertainty  is  a   consequence  of
no-signaling and  nonlocality.  As with Ineq.   (\ref{eq:Eta0}), Ineq.
(\ref{eq:U})  can  be  interpretted  as  showing  that  Bell-certified
uncertainty is weakened  by signaling. One can  still have uncertainty
originating from the separable component  of the correlations.  In QM,
this  would be  related to  quantum discord,  which arises  from local
non-commutativity, even for separable states \cite{OZ01}.

Our    result    (\ref{eq:ucmp})    also    generalizes,    for    the
two-input-two-output   case,   the   Oppenheim-Wehner   theorem   that
uncertainty bounds  nonlocality \cite{OW10},  which, in  our approach,
would be
\begin{equation}
\mathcal{U}_{\mathcal{A}} \ge \frac{C}{2},
\label{eq:OW}
\end{equation}  
obtained   from  Ineq.    (\ref{eq:U})  by   setting  $s:=0$.    Ineq.
(\ref{eq:ucmp}) can  be interpreted as asserting  that uncertainty and
signaling  jointly  upper-bound  nonlocality.   In  other  words,  the
nonlocality (as  quantified by $C$  rather than by the  probability to
win  a  CHSH   game  or  by  some  other  measure)   can  violate  the
Oppenheim-Wehner   uncertainty  bound   in   the  form   (\ref{eq:OW})
\textit{for the  case of signaling  correlations}.  In the  absence of
signaling,  uncertainty by  itself determines  nonlocality, and  Ineq.
(\ref{eq:U}) reduces to the result of \cite{OW10}.
 
It is worth noting our  uncertainty bound on nonlocality (\ref{eq:U}),
in contrast to  Ref.  \cite{OW10}, does not  invoke steering. Instead,
(\ref{eq:U})   can  be   seen   simply  as   a   consequence  of   the
complementarity  between   uncertainty  and  signaling   in  resources
required to  simulate nonseparable unpredictability, analogous  to the
complementarity  between local  randomness and  signaling to  simulate
singlet   statistics   \cite{ASQIC,KGB+11,Hal10b}.   The   reason   is
essentially   that  Ref.    \cite{OW10}   employs   the  approach   of
fine-grained uncertainty based on  a ``random access coding'' accessed
by Bob  conditioned on Alice's input  $a$ and output $A$,  which makes
the  uncertainty  ensemble-dependent,  whereas our  method  quantifies
uncertainty unilaterally,  as a ``conjugate  unpredictability'' rather
than by fine-graining.  This contrast can  be illustrated for a PR box
\cite{PR94}, which is maximally nonlocal  ($C=1$).  In the approach of
\cite{OW10}, a  PR box  lacks uncertainty,  and the  nonlocality comes
purely  from  perfect steering.   However,  according  to the  present
approach,  from  Ineq.   (\ref{eq:U}),   setting  $\eta=1$.   we  find
$\mathcal{U}=\frac{1}{2}$,  i.e.,   the  nonlocality   constrains  the
uncertainty to  be maximal.  For classical  systems, $\mathcal{U}$ and
$\eta$ identically vanish.

A  subtlety worth  noting here  is that  only in  the case  of spatial
correlations does  $\mathcal{U}$ truly represent uncertainty.   In the
temporal  case,  $\mathcal{U}$  must  be   interpreted  as  a  mix  of
measurement uncertainty and measurement disturbance \cite{ASX}.

\section{Conclusions}

Nonclassical  properties  like  intrinsic randomness,  no-cloning  and
uncertainty,  which  are   known  to  be  consequences   of  the  twin
assumptions of nonlocality and no-signaling, are shown to subsist even
when no-signaling  is relaxed, in the  context of two-input-two-output
correlations, provided there is a nonvanishing signal deficit, $\eta$,
which is  the excess of the  communication cost over the  signaling in
the correlation. This result, which  can also be generalized to higher
dimensions,  is   shown  to   imply  the  presence   of  non-seperable
unpredictability.  This forms our  criterion of strong nonclassicality
which is  independent of whether  the correlation is  spatial (between
two geographically separated particle) or temporal (between two events
on the  same particle). Weaker  versions of nonclassicality  were also
indicated.

In particular we show that signal diminishes strong nonclassicality in
the sense  that the  lower bound on  the nonclassical  properties like
unpredictability  and uncertainty  reduces with  increasing signal  at
fixed   communication  cost.   This   generalizes  to   the  case   of
non-vanishing signal the existence of unpredictability and uncertainty
etc proven for the nonlocal-nonsignaling correlations \cite{MAG06} and
the uncertainty bound on nonlocality \cite{OW10}.

\begin{acknowledgements}

We  thank   the  anonymous  referee   for  the  helpful   remarks  and
constructive suggestions.  SA acknowledges support through the INSPIRE
fellowship  [IF120025] by  the Department  of Science  and Technology,
Govt.  of  India, financial support  from IQSA, Adamar  Mutt education
foundation,  Mr.   Nandish  Jyothi  to  attend  IQSA  meeting  Quantum
2014.  SA   acknowledges  the  academic  assistance   of  the  Manipal
University graduate program. RS acknowledges  support from the DST for
project SR/S2/LOP-02/2012.

\end{acknowledgements}

%\bibliographystyle{spphys}
%\bibliography{quantarv}

\end{document}